# High performance distributed feedback quantum dot lasers with laterally coupled dielectric grating

Zhuohui Yang, Zhengqing Ding, Lin Liu, Hancheng Zhong, Sheng Cao, Xinzhong Zhang, Shizhe Lin, Xiaoying Huang, Huadi Deng, Ying Yu*, Siyuan Yu*

*State Key Laboratory of Optoelectronic Materials and Technologies, School of Electronics and Information Technology, Sun Yat-Sen University, Guangzhou 510275, China*

yuying26@mail.sysu.edu.cn

yusy@mail.sysu.edu.cn

**Abstract**

The combination of grating-based frequency-selective optical feedback mechanisms, such as distributed feedback (DFB) or distributed Bragg reflector (DBR) structures, with quantum dot (QD) gain materials is a main approach towards ultra-high-performance semiconductor lasers for many key novel applications, either as stand-alone sources or as on-chip sources in photonic integrated circuits. However, the fabrication of conventional buried Bragg grating structures on GaAs, GaAs/Si, GaSb and other material platforms have been met with major material regrowth difficulties. We report a novel and universal approach of introducing laterally coupled dielectric Bragg gratings to semiconductor lasers that allows highly controllable, reliable and strong coupling between the grating and the optical mode. We implement such a grating structure in a low-loss amorphous silicon material alongside GaAs lasers with InAs/GaAs QD gain layers. The resulting DFB laser arrays emit at pre-designed 0.8 THz LWDM frequency intervals in the 1300 nm band with record performance parameters, including side mode suppression ratios as high as 52.7 dB, continuous-wave output power of 27.7 mW (room-temperature) and 10 mW (at 70°C), and ultra-low relative intensity noise (RIN) of < -165 dB/Hz (2.5-25 GHz). The devices are also capable of operating isolator-free under very high external reflection levels of up to -12.3 dB whilst maintaining the high spectral and ultra-low RIN qualities. These results validate the novel laterally coupled dielectric grating as a technologically superior and potentially cost-effective approach for fabricating DFB and DBR lasers free of their semiconductor material constraints, thus universally applicable across different material platforms and wavelength bands.

## 1. Introduction

Embedding semiconductor lasers with Bragg gratings as the wavelength-selective feedback mechanism is a well-established approach to achieving high-quality single-frequency lasing. In conjunction with the distinctive properties of various compound semiconductor gain materials, distributed feedback (DFB) and distributed Bragg reflector (DBR) lasers are finding a wide range of applications in both classical and quantum domains, such as with InGaAs emitting in the near infrared for optical communication [1], with GaAs or GaAsP in the red spectral range [2] for atomic clocks [3], atom interferometry [4] and efficient optical pumping [5], with GaSb or InAs/AlSb in the middle to far infrared [6] for trace-gas sensing [7], and with nitride semiconductors in the green to ultraviolet for absorption spectroscopy [8] and high density data storage [9].

For optical data interconnect applications, the 1310 nm band is of particular interest for low-cost wavelength division multiplexed (WDM) system [10], 5G and 6G optical networks [11], as well as for LiDAR [12, 13] and sensing [14, 15]. Compared with conventional InGaAs/InGaAlAs quantum well (QW) materials emitting in the same wavelength range, self-assembled InAs/GaAs quantum-dots (QDs) have achieved superior performance, including lower threshold current [16] and higher temperature stability [17] due to the zero-dimensional (0-D) carrier confinement in the QDs. A highly desirable feature of QD lasers is their ultra-low relative intensity noise (RIN) [18-20] originating from their very low linewidth enhancement factor $\alpha$, which affords great advantages in analog transmission (such as Radio over Fiber, RoF) and sensing [21]. The low $\alpha$ also underpins their tolerance to high levels of external optical feedback [22-24], making QD lasers very promising light sources for reliable, scalable and isolator-free photonic integrated circuits (PICs) [25]. Furthermore, InAs/GaAs QD materials have also been proved to be highly tolerant to epitaxial defects [26-28], yielding high-performance Fabry-Perot (FP) type laser diodes epi-grown on silicon substrate [28-30]. Similar advantages have also been observed in other QD laser systems such as InAs/InP [31, 32].

For DFB lasers, the buried Bragg gratings, as first realized in InP-based 1550 nm lasers, are conventionally placed on top of the active layer and fabricated through a regrowth process after grating definition and etching. The proximity to the waveguide allows the grating to intercept the optical field with a high coupling coefficient $\kappa$. This conventional approach has also been attempted in other materials and wavelengths such as GaAs-based QD lasers. In 2011, Tanaka et al. demonstrated a 1293-nm InAs/GaAs QD DFB laser with a GaAs grating buried in a metal organic vapor phase epitaxy (MOVPE) regrown InGaP cladding, achieving a $\kappa$ of 40 cm$^{-1}$ and side-mode-suppression-ratio (SMSR) of 45 dB [33]. In 2018, Wan et al. demonstrated a 1310 nm QD DFB laser epitaxially on silicon by molecular beam epitaxy (MBE) regrowth. Using a GaAs grating buried in an Al$_{0.4}$Ga$_{0.6}$As upper cladding, a $\kappa$ of 45 cm$^{-1}$ and SMSR of >50 dB were achieved [34].

However, the regrowth process represents poor productivity for many material systems. For GaAs-based laser devices, an InGaP upper cladding layer requires wafer transfer from MBE to MOCVD system, while an AlGaAs upper cladding layer requires rigorous pretreatment before regrowth via ultrahigh-vacuum MBE chamber. Harder still, the regrowth process for GaN(Sb)-based lasers suffers from a lack of contamination-free AlGaN(Sb) regrowth process or other latticed-matched cladding materials with sufficient bandgap and refractive index contrasts.

An alternative, regrowth-free, approach uses Bragg gratings etched alongside a ridge waveguide to form a laterally coupled distributed-feedback (LC-DFB) laser. For InP-based laser structures, the gratings can be fabricated simultaneously during the waveguide etching process. Using an aluminium-containing stop-etch layer and a chemically selective recipe [35, 36], the grating penetration depth can be precisely delimited to just above the active layer to achieve a precise $\kappa$ value. However, translating this approach to the GaAs or GaSb based DFB laser structures has proven to be challenging due to the lack of a suitable selective etch-stop layer. Previously demonstrated GaAs [37], GaSb [38] or GaN [39] ridge waveguide with LC-gratings fabricated using such one-step reactive ion etching (RIE) approach suffer from the fact that, due to local chemical transportation and reaction rate variations caused by the etched ridge and the very narrow grating gaps, it is very difficult to control the etch depth at the foot of the ridge waveguide where the grating intercept the optical mode. An undesirable feature known as 'footing', which is a gradual increase of etch depth away from the foot of the etched waveguide, and another feature known as 'RIE-lag', which is a decreased etch depth in narrow gaps, result in significant uncertainties in the grating $\kappa$ value.

To circumvent this problem, in a previous work, the authors chose to etch the lateral grating deeply through the active region so that the grating etch depth no longer affects $\kappa$ [40]. However, a deep-etched active waveguide suffers from increased surface recombination and optical scattering loss, and a waveguide with practical widths can support more than one transverse mode, with the unwanted high-order modes as a favored lasing mode due to their higher $\kappa$ values [40]. For GaSb lasers, metal gratings deposited after waveguide etching [41] were also used. While providing strong optical coupling, metal gratings can introduce significant additional absorption loss in the laser cavity.

In this article, we demonstrate a novel dielectric grating structure placed alongside single transverse mode ridge waveguides that have a precisely controlled trapezoid cross-sectional profile etched to a depth just above the active layer. Fabricated in an amorphous silicon ($\alpha$-Si) layer deposited after the formation of the ridge waveguide, the grating corrugations, plasma-etched into the $\alpha$-Si, is precisely stopped at an underlying etch stop layer of $Al_2O_3$ deposited after the waveguide etching and before the $\alpha$-Si layer thanks to the chemical selectivity between the two materials. High contrast grating (with a refractive index difference of $\Delta n \sim 2$) is formed between the $\alpha$-Si corrugations and a subsequently deposited silicon dioxide ($SiO_2$) cladding layer, producing a LC grating with significantly enhanced and precisely controllable coupling coefficient $\kappa$.

We implemented the novel structure on an InAs/GaAs QD gain material, producing LC-DFB laser arrays emitting across the 1300 nm band on an 800 GHz local area network wavelength division multiplexing (LWDM) grid. The devices emit more than 27.7 mW of single-mode output per facet at room temperature and a typical SMSR greater than 52.7 dB. They also demonstrate ultra-low RIN of < -165 dB/Hz in the range of 2.5-25 GHz and isolator-free operation under external feedback levels of up to -12.3 dB (5.9%). The output power, SMSR and RIN values are the best of the reported values of InAs/GaAs QD LC-DFB lasers as far as we are aware. These superior performances of the devices validate the novel LC grating as an effective regrowth-free approach to grating-based laser fabrication. In addition to the elimination of regrowth, the deployment of the novel LC grating structure decouples its fabrication from specific laser materials, and therefore the scheme could serve as a universal alternative approach for high-performance semiconductor laser devices employing grating structures as optical feedback mechanisms.

## 2. Design and fabrication of the LC-DFB QD laser

The InAs/GaAs QD laser structure, in a typical p-i-n configuration, was grown on 3-inch semi-insulating GaAs (001) substrates in a solid-source MBE chamber. Details of material structure and characteristics are given in Section 1 of Supplement materials. As shown in Figure 1(a), a waveguide width of 2.1 µm and a depth of 1.7 µm are used to support only the lowest order transverse mode (TE00). The ridge waveguides were patterned using electron beam lithography (EBL) and etched using an optimized inductively coupled plasma reactive-ion etching (ICP-RIE) process, with a trapezoid cross-section, a sidewall slope of θ=76°, and near-zero 'footing' (Figure 1(b)). The optimization process is described in Section 2 of Supplement Materials. The near-ideal trapezoid waveguide profile is key to a deterministic grating coupling coefficient $\kappa$ and low scattering loss, both very important for improved DFB laser performance. In addition, a small footing gives rise to a larger $\kappa$, which results from the increased evanescent field into the grating region [42].

A 10 nm $Al_2O_3$ passivation layer and a 150-nm-thick α-Si (n~3.495) layer were deposited for grating fabrication, with both layers covering the entire sample surface terrain, including the sloped sidewalls. First-order gratings with a period Λ in a range of 194.5-199.7 nm, a grating duty ratio of 1:1 and an extrusion of 8 μm from the waveguide foot were designed and patterned by an EBL resist (ARP6200) alongside the ridge. A ¼-λ phase shift was placed in the middle of the gratings to force lasing in the defect mode. LWDM-compatible laser arrays were achieved by adjusting the grating period Λ of adjacent lasers on the same bar, with a change of ΔΛ=0.727 nm results in a wavelength incremental of 0.8 THz. The grating corrugations were etched through the α-Si using a fluoride-based RIE process, stopping at the $Al_2O_3$ layer. The SEM image of Figure 1(c) reveals a high-quality LC grating with very sharp and smoothing sidewalls. With near-zero footing and a naturally chemically selective etch recipe, this dielectric grating structure is less sensitive to processing variations and, thus, more manufacturable than a buried heterostructure DFB structure. The grating coupling coefficient $\kappa$ is calculated from the lateral electric field distribution and effective index of the fundamental transverse electric mode via coupled-mode theory [42-44], detailed in Section 3 of Supplement Materials. For the designed devices, a first-order grating with a duty ratio of 1:1 produces a calculated $\kappa$ of 3.24 $mm^{-1}$.

The corrugations were subsequently covered with a 200 nm layer of $SiO_2$ (n ~ 1.46) to form a high contrast grating prior to contact window opening and Ti/Au contact deposition. After being cleaved into bars, the two facets were respectively covered with a six-layer $SiO_2/TiO_2$ high-reflection (HR, 96.8%) costing and a one-pair $SiO_2/TiO_2$ anti-reflection (AR, 1.7%) coating to suppress the facet back-reflections and increase the output power (see Figure S4 of Supplement Materials). Finally, the laser arrays were mounted epitaxy-side up on gold-coated copper heat sinks.

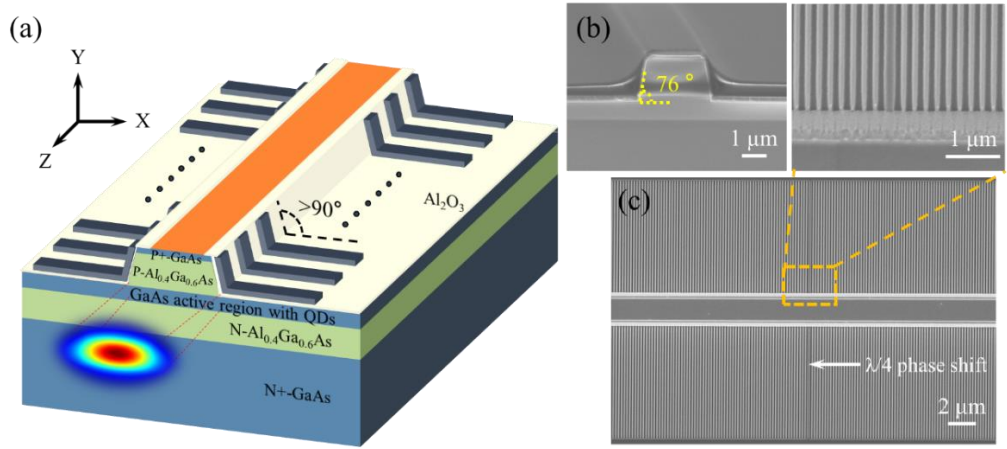

**Fig. 1.** (a) Schematic of the DFB laser structure, including the near-zero 'footing' trapezoid waveguide and the $\alpha$-Si gratings (not to scale); (b) Cross-sectional SEM image of the trapezoid waveguide with θ=76°, with the $\alpha$-Si and the ARP6200 photoresist layers also present; (c) SEM images of the etched $\alpha$-Si gratings with a $\lambda/4$ phase shift in the middle.

## 3. Lasing characteristics of the fabricated LC-DFB QD laser

At room temperature (25°C), a typical 2.1 μm × 1.9 mm DFB laser device (Figure 2(a)) has a turn-on voltage of 0.92 V and a differential series resistance of 11 Ω. The measured continuous wave (CW) threshold current of 19 mA corresponds to a current density of 476 A cm$^{-2}$. Above the threshold, the output power follows a kink-free near-linear curve with a slope efficiency of 0.21 W A$^{-1}$. The AR facet output power of 27.7 mW was obtained at an injection current of 150 mA or ~7.9×$I_{th}$. Figure 2(b) shows CW lasing up to 70 °C with output power of > 10 mW, and it is believed that the operation temperature can be further increased by more effective junction heat dissipation, either by further thinning the substrate or by flip-chip bonding.

The experimental value of $\kappa$ is estimated from the photonic bandgap width $\lambda_s$ = 0.436 nm observable from the below-threshold amplified spontaneous emission (ASE) spectrum of Figure 2(c). The total coupling strength $\kappa L$ for this 1.5 mm long device was estimated to be ~3.0 ($\kappa$=2.0), which is consistent with the theoretical value when taking into consideration the deviations of rating duty ratio (1:1.9 as shown inset of Figure 1(c)). When increasing the current to 80 mA (4.2×$I_{th}$), the central defect mode in seen in the ASE spectrum rises to be the dominant longitudinal mode with a SMSR of 52.7 dB (Figure 2(d)).

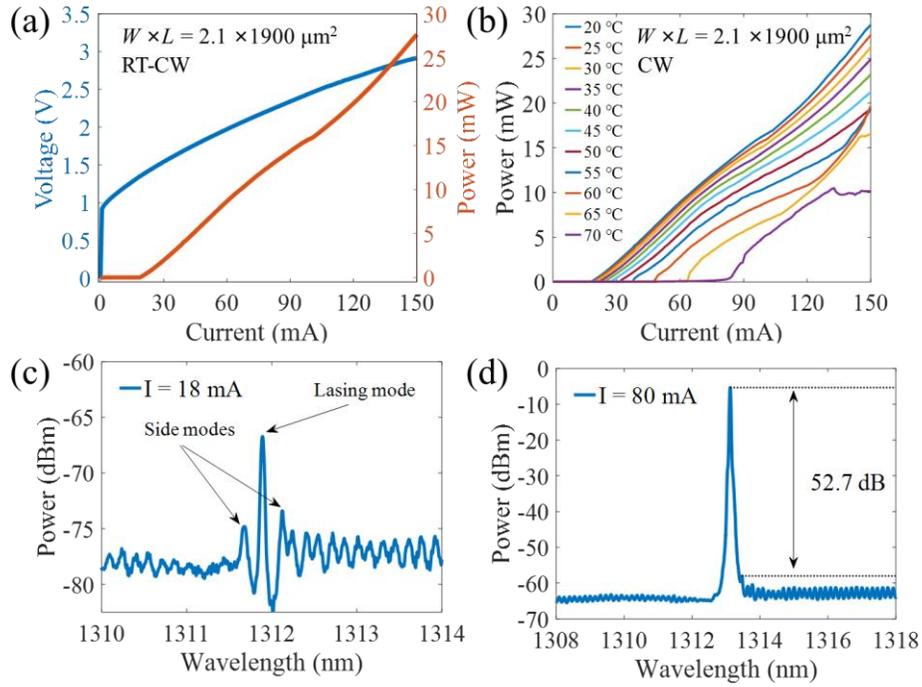

**Fig. 2.** (a) Typical L-I-V characteristics of a DFB laser with a 2.1×1900 μm² cavity at room temperature; (b) Temperature-dependence L-I curves from the DFB laser, showing lasing up to 70 °C under CW operation; (c-d) Optical spectra of a single DFB laser with a 2.1×1500 μm² cavity operating just below threshold (c) and at a drive current of 80 mA (d).

Stable single-mode operation was observed at CW currents up to 150 mA in the temperature ranging of 20 to 50°C, with a linear wavelength - temperature tuning rate of 0.12 nm/°C and a quadratic wavelength - current tuning curve (Figure 3 (a-b)). It is noteworthy that the laser maintains single-mode operation across the entire current range. This high single-mode quality is credited to the novel α-Si grating along the trapezoid waveguide which affords reliable deterministic optical coupling.

Across each eight-device bar, channel spacing of $0.80 \pm 0.10$ THz was measured, producing wavelengths ranging from 1300.05 nm to 1332.41 nm (Figure 3(c-d)), matching well with the standard LWDM grid. This wavelength range was limited by the range of grating periods we fabricated. The high precision electron beam lithography (EBL) process together with the broad gain bandwidth of the QD materials afford very promising applications for both LWDM and CWDM within the O-band.

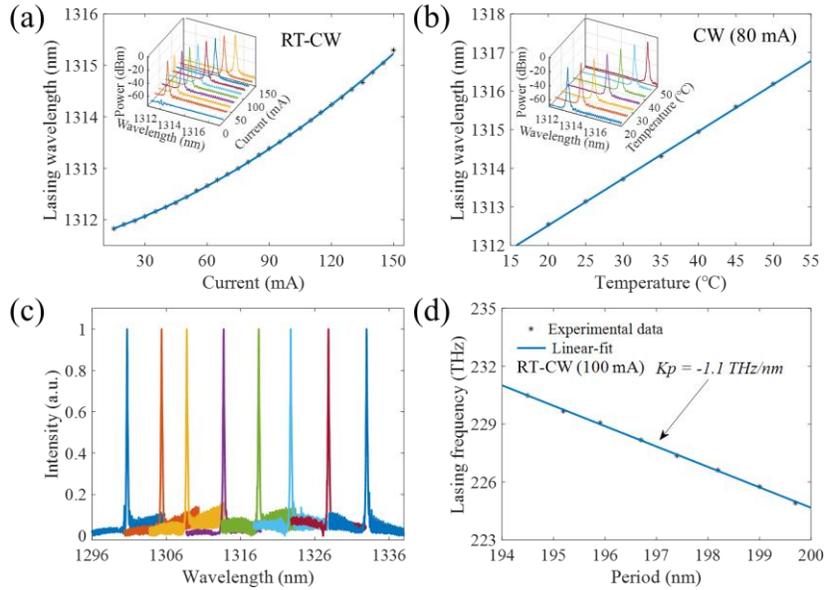

**Fig. 3.** (a) Wavelength shift with injection currents; (b) Wavelength shift with heat-sink temperature; (c-d) Optical spectra and lasing frequencies of a LWDM DFB laser array.

## 4. Relative intensity noise and external feedback sensitivity

The RIN across the frequency range of 2.5-20 GHz is assessed using a commercial system (SYCATUS A0010A), measuring < −155 dB/Hz at 77 mA ($4 \times I_{th}$) and reducing to a saturated minimum level of ∼ −165 dB/Hz at 172 mA ($9 \times I_{th}$), as shown in Figure 4. To the best of our knowledge, this ultra-low RIN is the best-reported result among QD DFB lasers and is in good agreement with reported values in both GaAs [20] and silicon [19] based QD-FP lasers.

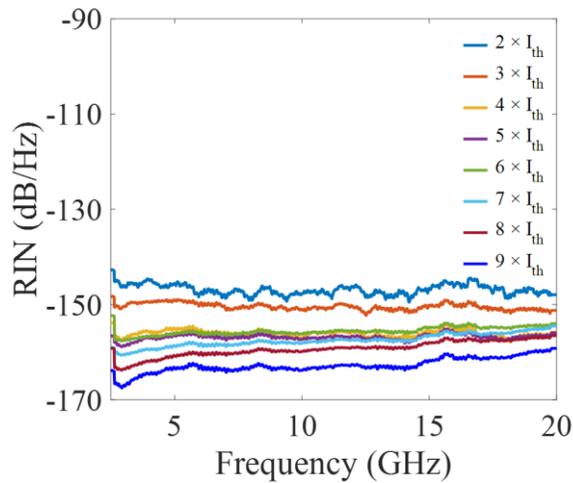

**Fig. 4.** Measured RIN spectra at several bias currents at 25°C.

Finally, the performance of the QD LC-DFB lasers under coherent optical feedback was assessed using the optical measurement setups shown schematically in Figure 5(a). The emission from the QD laser AR facet is coupled into the feedback test system by a lensed fiber with coupling efficiency of 20% - 30%, and divided into feedback and detection paths by a 90/10 fiber coupler, as illustrated in Figure S5. On the feedback path (10% of the coupled power), a fibered optical circulator is used to feed the light back to the laser cavity. Since the external cavity resonance frequency (17.25 MHz in this stage) is much less than the laser relaxation oscillation frequency (ROF), the impact of the feedback phase is negligible. The feedback strength $r_{ext}$ is defined as the ratio of the returning power to that of the laser free-space output power and the precise returning power can be obtained by calculating the product of the power (detected by power meter) and the lensed fiber coupling efficiency. While the optical feedback intensity is controlled by changing the operating current of a polarization-maintaining boost optical amplifier (BOA, Thorlabs S9FC1132P). A filter with 0.8 nm bandwidth is employed to spectrally suppress the amplified spontaneous emission from the BOA. A polarization controller is inserted in the external cavity to compensate for the polarization rotation in the fiber. The insertion losses produced in the lens fiber, BOA, beam splitter and each connector are carefully calibrated. The remaining 90% of the coupled power is sent to a high-resolution (0.03 nm) optical spectrum analyzer (OSA, Anritsu MS9740A) or RIN measurement system (SYCATUS A0010A) to monitor the evolution of spectra and RIN as the feedback strength $r_{ext}$ varies. For the whole measurement, the DFB laser is mounted on a thermo-electric cooler (TEC) operated at 25 °C.

Figures 5(b-c) presents the evolution of SMSR and RIN with the laser operating at 4×$I_{th}$. Little sign of deterioration is observed until the optical feedback levels reach $r_{ext}$ = 5.9 % (−12.3 dB). In particular, the SMSR of the laser is found to be still above 50 dB under this feedback strength. The RIN levels (at the frequency of 5 GHz) only rises slowly until $r_{ext}$ = 5.6% (−12.5 dB), without any visible periodic or chaotic oscillations in the RIN spectra. A sharp increase in the RIN indicates a transition to the coherence collapse regime beyond this critical level of optical feedback $f_{ext,c}$. For DFB laser, $f_{ext,c}$ is strongly associated with the normalized coupling coefficient $\kappa L$ ($L$ is the length of the laser cavity), where an increased $\kappa L$ will lead to an increased $f_{ext,c}$ [45, 46]. A linewidth enhancement factor $\alpha$ of 0.873 is derived from the measurement results (see detail parameters in Section 5 of Supplement Materials), which is much smaller than that of QW

lasers (normally in the range of 3-5) [47].

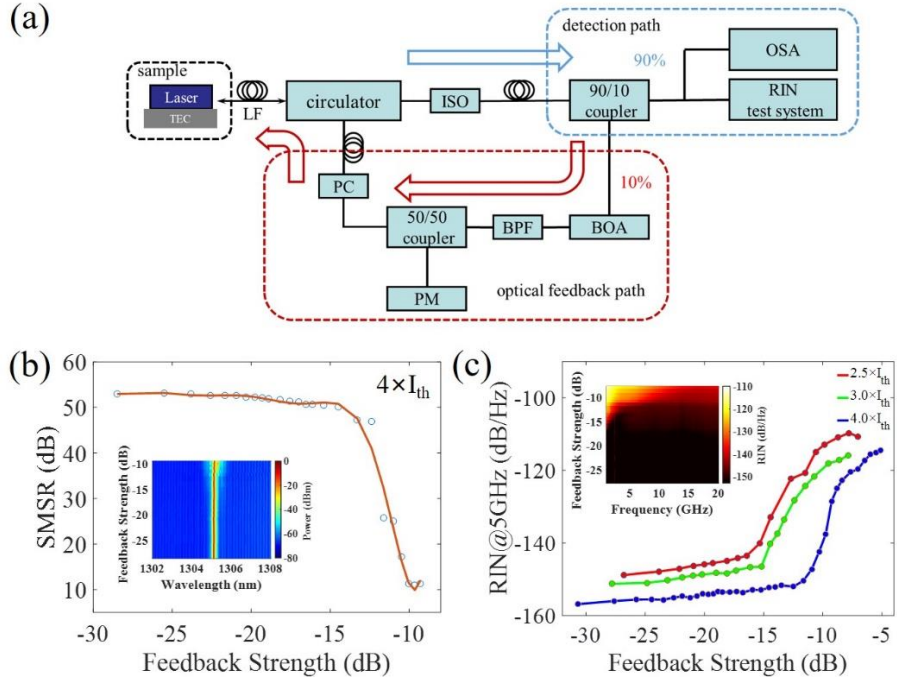

**Fig. 5.** (a) Experimental setup used for the long-delay feedback measurements (LF: lens fiber; PM: Power meter; BOA: boost optical amplifier; OSA: Optical Spectrum Analyzer; RIN: related intensity noise; PC: polarization controller; ISO: optical isolator; BPF: band-pass filter); (b) The evolution of the SMSR with the increasing feedback strength, the inset is the optical spectrum of the DFB laser as the feedback strength increases; (c) The change of the RIN in the same DFB laser under $2.5\times I_{th}$, $3\times I_{th}$ and $4\times I_{th}$ current injections. The inset is the frequency domain plot of the RIN as a function to the increasing feedback strength.

## 5. Discussion

Given the rapid development of QD-DFB laser at 1310 nm on both GaAs and silicon substrates, it is useful to compare the performance of our device with those reported in the literature. Table 1 lists the threshold current, power, highest work temperature, SMSR, RIN and optical feedback tolerance of reported DFB lasers together with our device. In general, compared with commercial QW lasers, the QD lasers exhibit excellent performance in terms of high operating temperature, low RIN and high optical feedback tolerance owing to their stronger carrier confinement, larger damping factor and smaller α factor. QD-laser with buried [19, 43] gratings show good feedback tolerance, but the RIN need to be further improved. Our device simultaneously achieves high output power (27.7 mW), ultra-low RIN (-165 dB/Hz) and high tolerance to optical feedback (-12.3 dB).

**Table 1.** Comparison of the performance of our device with the reference QD DFB laser at 1310 nm

| Year | Substrate | Grating | κ (mm⁻¹) | Threshold current (mA) | Power (mW) | SMSR (dB) | T (°C) | RIN (dB/Hz) | Anti-feedback (dB) | Ref. |
|---|---|---|---|---|---|---|---|---|---|---|
| 2003 | GaAs | GaAs sidewall | - | 3 | 9.3 | 50 | - | - | -14 | [48] |
| 2005 | GaAs | Metal sidewall | - | 5 | 14 | 50 | - | - | - | [49] |
| 2011 | GaAs | InGaP/GaAs buried | 4 | 6.8 | 10 | 45 | 80 | - | - | [33] |
| 2011 | GaAs | Cr sidewall | - | 18 | 20 | 53 | 85 | - | -12 | [50] |
| 2014 | GaAs | InGaP/GaAs buried | 2.5 | 43.8 | 34 | 58 | 60 | - | - | [51] |
| 2018 | GaAs | GaAs sidewall | - | 30 | 23 | 51 | - | - | - | [37] |
| 2018 | GaAs | InGaP/GaAs buried | 4 | 6.2 | 20 | 40 | 70 | -150 | -8 | [52] |
| 2018 | Heterogenously integrated GaAs/Si | Si | 7.7 | 9.5 | 2.5 | 47 | 100 | - | - | [53] |
| 2018 | Monolithic integrated GaAs/Si | GaAs sidewall | 4.2 | 12 | 1.5 | 50 | - | - | - | [40] |
| 2020 | Monolithic integrated GaAs/Si | GaAs sidewall | 4.5 | 20 | 4.4 | 50 | 70 | - | - | [34] |
| 2021 | GaAs | AlGaAs/GaAs buried | 1.6 | 9.3 | 15 | 50 | 55 | -150 | -6 | [24] |
| 2021 | GaAs | Amorphous Si sidewall | 2.0 | 19 | 27.7 | 52.7 | 70 | -165 | -12.3 | This work |

To conclude, by implementing a novel first-order amorphous silicon Bragg grating laterally coupled to a near-ideal trapezoid GaAs ridge waveguide, high-performance 1300-nm InAs/GaAs quantum dot DFB laser arrays have been realized with high power, ultra-low RIN, high robustness against optical feedback and accurate LWDM grid. These excellent features make the device a very attractive candidate for high performance digital and analog WDM optical transmission systems as well as on-chip sources for PICs where optical isolators are not readily available.

The novel grating structure affords accurate and deterministic grating coupling coefficient $\kappa$ that can be engineered independent of the laser material epitaxy process. The scheme can therefore be readily implemented on other material systems such as InP-, GaAs/Si- and GaSb-based compound semiconductor lasers, and in other wavelength windows by scaling the size of the grating and using low absorption dielectric materials for those wavelengths. The simplicity and versatility of the regrowth-free scheme makes it possible to establish a new paradigm of semiconductor laser manufacturing in which the co-manufacturing, on the same fabrication platform, of multiple types of grating-based semiconductor lasers (including DFB, DBR and tunable lasers) very different in their active materials and operating wavelengths would potentially reduce their manufacturing cost very significantly.

**Funding**
This work is supported by the National Key R&D Program of China (2018YFB2200201), the Science and Technology Program of Guangzhou (202103030001), the National Key R&D Program of Guang-dong Province (2020B0303020001), the Science Foundation of Guangzhou City for the Pearl River Star (201906010090), and the Local Innovative and Research Teams Project of Guangdong Pearl River Talents Program (2017BT01121).

**Disclosures**
The authors declare no conflict of interest.